\documentclass[%
 reprint,
nofootinbib,
 amsmath,amssymb,
 aps,
prd,
]{revtex4-1}

\usepackage{graphicx}% Include figure files
\usepackage{dcolumn}% Align table columns on decimal point
\usepackage{bm}% bold math
\usepackage{hyperref}% add hypertext capabilities
\usepackage{float}
\begin{document}

\preprint{APS/123-QED}

\title{Post-Limber Weak Lensing Bispectrum, Reduced Shear Correction, and Magnification Bias Correction}% Force line breaks with \\

\author{Anurag C. Deshpande}
\email{anurag.deshpande.18@ucl.ac.uk}
\author{Thomas D. Kitching}%
\affiliation{%
Mullard Space Science Laboratory, University College London, Holmbury St. Mary, Dorking, Surrey, RH5 6NT, UK}%

\date{\today}% It is always \today, today,

\begin{abstract}
The significant increase in precision that will be achieved by Stage IV cosmic shear surveys means that several currently used theoretical approximations may cease to be valid. An additional layer of complexity arises from the fact that many of these approximations are interdependent; the procedure to correct for one involves making another. Two such approximations that must be relaxed for upcoming experiments are the reduced shear approximation and the effect of neglecting magnification bias. Accomplishing this involves the calculation of the convergence bispectrum; typically subject to the Limber approximation. In this work, we compute the post-Limber convergence bispectrum, and the post-Limber reduced shear and magnification bias corrections to the angular power spectrum for a \emph{Euclid}-like survey. We find that the Limber approximation significantly overestimates the bispectrum when any side of the bispectrum triangle, $\ell_i<60$. However, the resulting changes in the reduced shear and magnification bias corrections are well below the sample variance for $\ell\leq5000$. We also compute a worst-case scenario for the additional biases on $w_0w_a$CDM cosmological parameters that result from the difference between the post-Limber and Limber approximated forms of the corrections. These further demonstrate that the reduced shear and magnification bias corrections can safely be treated under the Limber approximation for upcoming surveys.
\end{abstract}

\maketitle

%\tableofcontents

\section{\label{sec:intro}Introduction}

Weak gravitational lensing can be a powerful tool to better constrain our knowledge of the currently favoured standard model for the Universe, the Lambda Cold Dark Matter model ($\Lambda$CDM). A useful manifestation of this effect is cosmic shear: the distortion of the observed shapes of distant galaxies due to weak gravitational lensing by the large-scale structure of the Universe (LSS). By measuring this distortion in large samples of galaxies, we can probe the LSS. Since the development of this structure depends on the fundamental properties of the Universe, measuring these distortions allows us to constrain cosmological parameters. Using the technique of tomography (where the observed population of galaxies is divided into different redshift bins), three-dimensional information can be obtained. In particular, cosmic shear can put strong constraints on the dark energy \citep{DETFrep}.

Contemporary cosmic shear surveys \cite{cfhtmain, DESpap, kids450} are able to carry out precision cosmology competitive with recent probes of the Cosmic Microwave Background \cite{Planck18}. Additionally, impending Stage IV \cite{DETFrep} weak lensing experiments such as \emph{Euclid}\footnote{\url{https://www.euclid-ec.org/}} \cite{EuclidRB}, \emph{WFIRST}\footnote{\url{https://www.nasa.gov/wfirst}} \cite{WFIRSTpap}, and LSST\footnote{\url{https://www.lsst.org/}} \cite{LSSTpap}, will have over an order-of-magnitude more precision than existing surveys \cite{SellentinStarck19}.

This presents a challenge: approximations made in our theoretical analyses may no longer be valid. Accordingly, a thorough examination of these effects is necessary. One such approximation, that is regularly made, is the Limber approximation. In this, only wave-modes in the plane of the sky are considered to be contributing to the lensing signal. The impact of relaxing this approximation, together with the the Hankel transform and flat-sky approximations, for a \emph{Euclid}-like experiment has been evaluated \cite{limitsofshear17}. Two further effects that have recently shown to be important for Stage IV experiments are the reduced shear approximation and magnification bias \cite{Deshpap}. Compounding the complexity of correcting for any one such approximation is that the procedure for doing so often involves making one of the others.

In this work, we focus on the reduced shear approximation and magnification bias, and their inter-dependency with the Limber approximation. When cosmic shear is probed, the quantity measured is reduced shear, rather than shear itself. Under the reduced shear approximation, the statistics of one are taken to equal those of the other. On the other hand, magnification bias refers to the change of the observed galaxy number density due to individual sources or patches of the sky being magnified. Magnification bias also affects probes of galaxy clustering \cite{DuncMag, Lormag, thielmag}. These two effects are treated together because their corrections take mathematically similar forms \cite{RSMBcombpap}. However, these corrections depend on the convergence bispectrum, and typically evaluate this quantity under the Limber approximation. 

Here, we forgo the Limber approximation when calculating the convergence bispectrum. Subsequently, we measure the resulting change in the magnification bias and reduced shear corrections for a \emph{Euclid}-like survey, and compare it to the sample variance of the survey. We also demonstrate that the resulting change does not induce significant biases in inferred $w_0w_a$CDM cosmological parameters if neglected.

This paper is organised as follows: In Section \ref{sec:theory}, we present the theoretical formalism. We begin by reviewing the basic cosmic shear power spectrum calculation. We also describe two additional components of the observed shear power spectrum: non-cosmological signals from the intrinsic alignments (IA) of galaxies, and shot noise. Next, the reduced shear and magnification bias calculations are reviewed. We then discuss the convergence bispectrum with and without the Limber approximation. Additionally, we explain the formalism we use to predict any resulting biases in inferred cosmological parameters. Following this, in Section \ref{sec:method}, explain our methodology; discussing modeling specifics and our choice of fiducial cosmology. Finally, in Section \ref{sec:results}, we present our results and discuss their consequences for Stage IV experiments. We first discuss the impact of the Limber approximation on the convergence bispectrum, and then the resulting impact on the reduced shear and magnification bias corrections.

\section{\label{sec:theory}Theory}

We begin by reviewing the first-order calculation of the cosmic shear angular power spectrum. Next, we describe the corresponding corrections for the reduced shear approximation and magnification bias, and their dependence on the convergence bispectrum. Finally, the calculation of the convergence bispectrum, with and without the Limber approximation, is explained.

\subsection{\label{subsec:angpowfo}The First-order Cosmic Shear Power Spectrum}

The change in the observed ellipticity of a distant galaxy due to weak lensing by the LSS is dependent on the reduced shear, $g$:
\begin{equation}
    \label{eq:redshear}
    g^\alpha(\boldsymbol{\theta})= \frac{\gamma^\alpha(\boldsymbol{\theta})}{1-\kappa(\boldsymbol{\theta})},
\end{equation}
where $\boldsymbol{\theta}$ is the source's position on the sky, $\gamma$ is the shear, and $\kappa$ is the convergence. These two terms encode the two different types of distortion from weak lensing; shear is the anisotropic stretching that makes circular distributions of light elliptical, and convergence is the isotropic increase or decrease in the size of the image. The index $\alpha$ encodes the fact that the shear is a spin-2 quantity. Now, in the weak lensing regime, $|\kappa|\ll 1$ so the reduced shear approximation is typically made for equation (\ref{eq:redshear}):
\begin{equation}
    \label{eq:RSA}
    g^\alpha(\boldsymbol{\theta}) \approx \gamma^\alpha(\boldsymbol{\theta}).
\end{equation}

The convergence of a galaxy image in tomographic redshift bin, $i$, is given by the projection of the density contrast of the Universe, $\delta$, along the line-of-sight over comoving distance, $\chi$, to the survey's limiting comoving distance, $\chi_{\rm lim}$:
\begin{equation}
    \label{eq:convergence}
    \kappa_i(\boldsymbol{\theta})=\int_{0}^{\chi_{\rm lim}} {\rm d}\chi\:\delta[S_{\rm K}(\chi)\boldsymbol{\theta},\, \chi]\:W_i(\chi).
\end{equation}
Here, $S_{\rm K}(\chi)$ is a function which accounts for the curvature of the Universe, $K$:
\begin{equation}
    \label{eq:SK}
    S_{\rm K}(\chi) = \begin{cases}
    |K|^{-1/2}\sin(|K|^{-1/2}\chi) & \text{\small{$K>0$ (Closed)}}\\
    \chi & \text{\small{$K=0$ (Flat)}}\\
    |K|^{-1/2}\sinh(|K|^{-1/2}\chi) & \text{\small{$K<0$ (Open)}.}
  \end{cases}
\end{equation}
Also, $W_i$ is the lensing kernel for bin $i$ \cite{Shapiro09}:
\begin{align}
    \label{eq:Wi}
    W_i(\chi) &= \frac{3}{2}\Omega_{\rm m}\frac{H_0^2}{c^2}\frac{S_{\rm K}(\chi)}{a(\chi)}\int_{\chi}^{\chi_{\rm lim}}{\rm d}\chi'\:n_i(\chi')\nonumber\\
    &\times\frac{S_{\rm K}(\chi'-\chi)}{S_{\rm K}(\chi')},
\end{align}
where $\Omega_{\rm m}$ is the dimensionless present-day matter density parameter of the Universe, $H_0$ is the Hubble constant, $c$ is the speed of light in a vacuum, $a(\chi)$ is the scale factor of the Universe, and $n_i(\chi)$ is the probability distribution of galaxies within bin $i$.

We can then relate the spin-2 shear to the convergence, in spherical harmonic space, with:
\begin{equation}
    \label{eq:fourier}
    \widetilde{\gamma}_i^\alpha(\boldsymbol{\ell})= T^\alpha(\boldsymbol{\ell})\,\widetilde{\kappa}_i(\boldsymbol{\ell}),
\end{equation}
where $\boldsymbol{\ell}$ is the spherical harmonic conjugate of $\boldsymbol{\theta}$, the `prefactor unity' approximation \cite{limitsofshear17} has been made, and $T^\alpha(\boldsymbol{\ell})$ are trigonometric weighting functions with the definitions:
\begin{align}
    \label{eq:Trigfunc1}
    T^1(\boldsymbol{\ell}) &= \cos(2\phi_\ell),\\
    \label{eq:Trigfunc2}
    T^2(\boldsymbol{\ell}) &= \sin(2\phi_\ell),
\end{align}
where $\phi_\ell$ is the angular component of vector $\boldsymbol{\ell}$ with magnitude $\ell$.

For an arbitrary shear field, we can construct two linear combinations of the shear components: a curl-free $E$-mode, and a divergence-free $B$-mode:
\begin{align}
    \label{eq:Emode}
    \widetilde{E}_i(\boldsymbol{\ell})&=\sum_\alpha T^\alpha\:\widetilde{\gamma}_i^\alpha(\boldsymbol{\ell}),\\
    \label{eq:Bmode}
    \widetilde{B}_i(\boldsymbol{\ell})&=\sum_\alpha \sum_\beta \varepsilon^{\alpha\beta}\,T^\alpha(\boldsymbol{\ell})\:\widetilde{\gamma}_i^\beta(\boldsymbol{\ell}),
\end{align}
in which $\varepsilon^{\alpha\beta}$ is the two-dimensional Levi-Civita symbol, where $\varepsilon^{12}=-\varepsilon^{21}=1$ and $\varepsilon^{11}=\varepsilon^{22}=0$. In the absence of higher-order systematic effects, the $B$-mode then vanishes. We are left with the $E$-mode, for which we can define auto and cross-correlation power spectra, $C_{\ell;ij}^{\gamma\gamma}$:
\begin{equation}
    \label{eq:powerspecdef}
    \left<\widetilde{E}_i(\boldsymbol{\ell})\widetilde{E}_j(\boldsymbol{\ell'})\right> = (2\pi)^2\,\delta_{\rm D}^2(\boldsymbol{\ell}+\boldsymbol{\ell'})\,C_{\ell;ij}^{\gamma\gamma},
\end{equation}
where $\delta_{\rm D}^2$ is the two-dimensional Dirac delta. Under the Limber approximation, where it is assumed that only $\ell$-modes in the plane of the sky contribute to the lensing signal, the power spectra themselves can then be written as:
\begin{equation}
    \label{eq:Cl}
    C_{\ell;ij}^{\gamma\gamma} = \int_0^{\chi_{\rm lim}}{\rm d}\chi\frac{W_i(\chi)W_j(\chi)}{S^{\,2}_{\rm K}(\chi)}P_{\delta\delta}(k, \chi),
\end{equation}
where $P_{\delta\delta}(k, \chi)$ is the matter overdensity power spectrum. Comprehensive reviews of this standard calculation can be found in \cite{Kilbinger15, Munshirev}.

\subsection{\label{subsec:IAs}Intrinsic Alignments}

In fact, the angular power spectrum we measure from the ellipticities of galaxies contains not only the cosmic shear contribution of equation (\ref{eq:Cl}) but also other, non-cosmological parts. One such contribution comes from the intrinsic alignment (IA) of galaxies \cite{JoachimiIAs}. 

Galaxies that form near each other do so in similar tidal environments. This causes those galaxies to have preferred, intrinsically correlated, alignments. The ellipticity of a galaxy, $\epsilon$ can be described to first-order as:
\begin{equation}
    \label{eq:galelip}
    \epsilon = \gamma + \gamma^{\rm I} + \epsilon^s,
\end{equation}
where $\gamma=\gamma^1+{\rm i}\gamma^2$ is the cosmic shear term, $\gamma^{\rm I}$ is the contribution from IAs, and $\epsilon^s$ is the galaxy's source ellipticity in the absence of any IA. The theoretical two-point statistic (e.g. the power spectrum) calculated from equation (\ref{eq:galelip}) consists of three types of terms: $\langle\gamma\gamma\rangle,\langle\gamma^{\rm I}\gamma\rangle$, and $\langle\gamma^{\rm I}\gamma^{\rm I}\rangle$.

The first of these terms corresponds to the cosmic shear power spectra from equation (\ref{eq:Cl}). Meanwhile the other two terms result in additional contributions to the observed power spectra, $C_{\ell;ij}^{\epsilon\epsilon}$, so that:
\begin{equation}
    \label{eq:ObsCl}
    C_{\ell;ij}^{\epsilon\epsilon} = C_{\ell;ij}^{\gamma\gamma} + C_{\ell;ij}^{{\rm I}\gamma} + C_{\ell;ij}^{\gamma{\rm I}} + C_{\ell;ij}^{\rm II} + N_{\ell;ij}^\epsilon,
\end{equation}
where $C_{\ell;ij}^{{\rm I}\gamma}$ represents the correlation between the background shear and the foreground IA, $C_{\ell;ij}^{\gamma{\rm I}}$ is the correlation of the foreground shear with background IA and is zero except in the case of photometric redshifts causing scattering of observed redshifts between bins, $C_{\ell;ij}^{\rm II}$ is the auto-correlation spectra of the IAs, and $N_{\ell;ij}^\epsilon$ is a shot noise term.

The additional spectra can be described analogously to the shear power spectra, by way of the non-linear alignment (NLA) model \citep{NLAmodel}:
\begin{align}
    \label{eq:cllig}
    C_{\ell;ij}^{{\rm I}\gamma} &= \frac{(\ell+2)!}{(\ell-2)!}\frac{1}{(\ell+1/2)^4}\int_0^{\chi_{\rm lim}}\frac{{\rm d}\chi}{S^{\,2}_{\rm K}(\chi)}[W_i(\chi)n_j(\chi) \nonumber\\
    &+n_i(\chi)W_j(\chi)] P_{\delta {\rm I}}(k, \chi),\\
    \label{eq:clli}
    C_{\ell;ij}^{\rm II} &= \frac{(\ell+2)!}{(\ell-2)!}\frac{1}{(\ell+1/2)^4}\int_0^{\chi_{\rm lim}}\frac{{\rm d}\chi}{S^{\,2}_{\rm K}(\chi)}n_i(\chi)\nonumber\\
    &\times n_j(\chi)\,P_{\rm II}(k, \chi),
\end{align}
where the intrinsic alignment power spectra, $P_{\delta {\rm I}}(k, \chi)$ and $P_{\rm II}(k, \chi)$, can be expressed as functions of the matter power spectra:
\begin{align}
    \label{eq:pdi}
    P_{\delta {\rm I}}(k, \chi) &= \bigg(-\frac{\mathcal{A}_{\rm IA}\mathcal{C}_{\rm IA}\Omega_{\rm m}}{D(\chi)}\bigg)\:\:P_{\delta\delta}(k,\chi),\\
    \label{eq:pii}
    P_{\rm II}(k, \chi) &= \bigg(-\frac{\mathcal{A}_{\rm IA}\mathcal{C}_{\rm IA}\Omega_{\rm m}}{D(\chi)}\bigg)^2P_{\delta\delta}(k,\chi).
\end{align}
Here, $\mathcal{A}_{\rm IA}$ and $\mathcal{C}_{\rm IA}$ are free model parameters to be determined by fitting to data or simulations, and $D(\chi)$ is the growth factor of density perturbations in the Universe, as a function of comoving distance.

\subsection{\label{subsec:shotnoise}Shot noise}

The final term in equation (\ref{eq:ObsCl}) is the result of the uncorrelated part of the unlensed source ellipticities; represented by $\epsilon^s$ in equation (\ref{eq:galelip}). For cross-correlation spectra this term is zero, because the ellipticities of galaxies at different comoving distances should be uncorrelated. However, for auto-correlation spectra, assuming that the tomographic bins in the survey are equi-populated, it is written as:
\begin{equation}
    \label{eq:shotnoise}
    N_{\ell;ij}^\epsilon = \frac{\sigma_\epsilon^2}{\bar{n}_{\rm g}/N_{\rm bin}}\delta_{ij}^{\rm K},
\end{equation}
where $\sigma_\epsilon^2$ is the variance of the observed ellipticities in the galaxy sample, $\bar{n}_{\rm g}$ is the galaxy surface density of the survey, $N_{\rm bin}$ is the number of tomographic bins used, and $\delta_{ij}^{\rm K}$ is the Kronecker delta.

\subsection{\label{subsec:RScor}The Reduced Shear Correction}

The reduced shear approximation can be relaxed by Taylor expanding equation (\ref{eq:redshear}) around $\kappa=0$, and retaining terms up to and including second-order so that equation (\ref{eq:RSA}) becomes \cite{Deshpap, KrauseHirata, Shapiro09}:
\begin{equation}
    \label{eq:gexpan}
    g^\alpha(\boldsymbol{\theta})=\gamma^\alpha(\boldsymbol{\theta})+(\gamma^\alpha\kappa)(\boldsymbol{\theta})+\mathcal{O}(\kappa^3).
\end{equation}
By substituting $g^\alpha$, as defined by equation (\ref{eq:gexpan}), for $\gamma^\alpha$ in equation (\ref{eq:Emode}) and recomputing, we recover equation (\ref{eq:powerspecdef}) plus a second-order correction term:
\begin{align}
    \label{eq:ecorr}
    \delta\left<\widetilde{E}_i(\boldsymbol{\ell})\widetilde{E}_j(\boldsymbol{\ell'})\right> &= (2\pi)^2\,\delta_{\rm D}^2(\boldsymbol{\ell}+\boldsymbol{\ell'})\:\delta C^{\rm RS}_{\ell;ij} \nonumber\\ &=  \sum_\alpha \sum_\beta T^\alpha(\boldsymbol{\ell})T^\beta(\boldsymbol{\ell'})\nonumber\\
    &\times\left<\widetilde{(\gamma^\alpha\kappa)}_i(\boldsymbol{\ell})\:\widetilde{\gamma}_j^\beta(\boldsymbol{\ell'})\right> \nonumber\\
    &+ T^\alpha(\boldsymbol{\ell'})T^\beta(\boldsymbol{\ell})\nonumber\\
    &\times\left<\widetilde{(\gamma^\alpha\kappa)}_j(\boldsymbol{\ell'})\:\widetilde{\gamma}_i^\beta(\boldsymbol{\ell})\right>,
\end{align}
where $\delta C^{\rm RS}_{\ell;ij}$ is then the reduced shear correction to the angular power spectra. This takes the form:
\begin{align}
    \label{eq:dCl}
    \delta C^{\rm RS}_{\ell;ij} &= \int_0^\infty\frac{{\rm d}^2\boldsymbol{\ell'}}{(2\pi)^2}\cos(2\phi_{\ell'}-2\phi_\ell) \nonumber\\
    &\times \left[B_{iij}^{\kappa\kappa\kappa}(\boldsymbol{\ell_1}, \boldsymbol{\ell_2}, \boldsymbol{\ell_3}) + B_{ijj}^{\kappa\kappa\kappa}(\boldsymbol{\ell_1}, \boldsymbol{\ell_2}, \boldsymbol{\ell_3})\right],
\end{align}
where $B_{iij}^{\kappa\kappa\kappa}$ and $B_{ijj}^{\kappa\kappa\kappa}$ are the three-redshift convergence bispectra. Due to the assumption of isotropy in the Universe, we are free to choose $\phi_{\ell}=0$.

\subsection{\label{subsec:MBcor}The Magnification Bias Correction}

The density of galaxies observed by a survey is also altered by gravitational lensing \cite{MBorig}. This effect manifests in two competing ways. In one, individual galaxies are magnified; causing their fluxes to be increased. Given that any galaxy survey will have some flux limit, this can cause sources that should otherwise be excluded having their flux increased enough to be included in the sample. On the other hand, the patch of sky around such sources will also be magnified. This would result in the galaxy density in that patch of sky being diluted. The net effect on the number density is known as magnification bias, and depends on the slope of the intrinsic, unlensed, galaxy luminosity function, at the survey’s flux limit.

In the weak lensing regime, the observed galaxy overdensity accounting for magnification bias, in tomographic bin $i$, can be expressed as \cite{MBHui, MBorig}:
\begin{equation}
    \label{eq:galover}
    \delta^g_{{\rm obs}; i}(\boldsymbol{\theta}) = \delta^g_i(\boldsymbol{\theta}) + (5s_i-2)\kappa_i(\boldsymbol{\theta}),
\end{equation}
where $\delta^g_i(\boldsymbol{\theta})$ is the intrinsic, unlensed, galaxy overdensity in bin $i$, and $s_i$ is the slope of the cumulative galaxy number counts brighter than the survey's limiting magnitude, $m_{\rm lim}$, for the redshift bin $i$. Here, we have assumed that fluctuations in the intrinsic galaxy overdensity are small on the scales of interest. The slope of the luminosity function is:
\begin{equation}
    \label{eq:slope}
    s_i = \frac{\partial{\rm log}_{10}\,\mathfrak{n}(\bar{z_i}, m)}{\partial m}\bigg|_{m_{\rm lim}},
\end{equation}
where $\mathfrak{n}(\bar{z_i}, m)$ is the true distribution of galaxies, evaluated at the central redshift of bin $i$, $\bar{z_i}$.

In practice, accounting for magnification bias is equivalent to replacing the true shear, $\gamma^\alpha_i$, by an `observed' shear, within the estimator used to determine the angular power spectrum from data:
\begin{align}
    \label{eq:MBshear}
    \gamma^\alpha_{{\rm obs}; i}(\boldsymbol{\theta}) \xrightarrow{} \gamma^\alpha_i(\boldsymbol{\theta})+\gamma^\alpha_i(\boldsymbol{\theta})\delta^g_{{\rm obs}; i}(\boldsymbol{\theta}) &= \gamma^\alpha_i(\boldsymbol{\theta})\nonumber\\
    &+\gamma^\alpha_i(\boldsymbol{\theta})\delta^g_i(\boldsymbol{\theta}) \nonumber\\
    &+ (5s_i-2)\nonumber\\
    &\times\gamma^\alpha_i(\boldsymbol{\theta})\kappa_i(\boldsymbol{\theta}).
\end{align}
Analogously to the procedure for reduced shear, we now substitute equation (\ref{eq:MBshear}) into equation (\ref{eq:Emode}) and recompute the E-mode product average. Source-lens clustering terms are negligible \cite{RSMBcombpap}, so we recover equation (\ref{eq:powerspecdef}), and an additional correction term:
\begin{align}
    \label{eq:MBEcorr}
    \delta\left<\widetilde{E}_i(\boldsymbol{\ell})\widetilde{E}_j(\boldsymbol{\ell'})\right> &= (2\pi)^2\,\delta_{\rm D}^2(\boldsymbol{\ell}+\boldsymbol{\ell'})\:\delta C^{\rm MB}_{\ell;ij} \nonumber\\
    &= \sum_\alpha \sum_\beta T^\alpha(\boldsymbol{\ell})T^\beta(\boldsymbol{\ell'})(5s_i-2)\nonumber\\
    &\times\left<\widetilde{(\gamma^\alpha\kappa)}_i(\boldsymbol{\ell})\,\widetilde{\gamma}_j^\beta(\boldsymbol{\ell'})\right>\nonumber\\
    &+ T^\alpha(\boldsymbol{\ell'})T^\beta(\boldsymbol{\ell})(5s_j-2)\nonumber\\
    &\times\left<\widetilde{(\gamma^\alpha\kappa)}_j(\boldsymbol{\ell'})\,\widetilde{\gamma}_i^\beta(\boldsymbol{\ell})\right>.
\end{align}
The corresponding correction to the angular power spectra, $\delta C^{\rm MB}_{\ell;ij}$, is:
\begin{align}
    \label{eq:dClMB}
    \delta C^{\rm MB}_{\ell;ij} &= \int_0^\infty\frac{{\rm d}^2\boldsymbol{\ell'}}{(2\pi)^2}\cos(2\phi_{\ell'}-2\phi_\ell)\nonumber\\
    &\times [(5s_i-2)B_{iij}^{\kappa\kappa\kappa}(\boldsymbol{\ell}, \boldsymbol{\ell'}, -\boldsymbol{\ell}-\boldsymbol{\ell'})\nonumber\\
    &+ (5s_j-2)B_{ijj}^{\kappa\kappa\kappa}(\boldsymbol{\ell}, \boldsymbol{\ell'}, -\boldsymbol{\ell}-\boldsymbol{\ell'})],
\end{align}
which is equivalent to equation (\ref{eq:dCl}) with factors of $(5s_i - 2)$ and $(5s_j - 2)$ applied to their respective bispectra contributions. Accordingly, these effects can both be computed for the computational cost of computing one.

\subsection{\label{subsec:bispec}The Convergence Bispectrum}

The corrections encompassed by equation (\ref{eq:dCl}) and equation (\ref{eq:dClMB}) both rely on calculating the convergence bispectrum. In its most general form, the observed convergence in spherical harmonic space, on a sphere is:
\begin{align}
    \label{eq:convspehe}
    \widetilde{\kappa}_{i; \ell m} &= 4\pi i^\ell \int_0^{\chi_{\rm lim}} {\rm d}\chi W_i(\chi) \int_0^\infty \frac{{\rm d}^3 k}{(2\pi)^3}j_\ell(k\chi)\nonumber\\
    &\times {}_2Y^*_{\ell m}(\boldsymbol{\hat{k}})\widetilde{\delta}(\boldsymbol{k}, \chi),
\end{align}
where $j_\ell$ are spherical Bessel functions, ${}_2Y^*_{\ell m}$ are spin-weighted with spin$=2$ spherical harmonics, $\widetilde{\delta}$ is the density contrast of the Universe in spherical harmonic space, and $\boldsymbol{k}$ is a spatial momentum vector with magnitude $k=|\boldsymbol{k}|$. 

Then, the bispectrum is the three-point counterpart of the power spectrum, and is defined on the sphere as \cite{IASpap}:
\begin{align}
    \label{eq:bidef}
     \left<\widetilde{\kappa}_{i;\ell_1m_1}\widetilde{\kappa}_{j;\ell_2m_2}\widetilde{\kappa}_{q;\ell_3m_3}\right> &= \mathcal{G}^{\ell_1 \ell_2 \ell_3}_{m_1m_2m_3}\nonumber\\
     &\times B_{ijq}^{\kappa\kappa\kappa}(\boldsymbol{\ell_1}, \boldsymbol{\ell_2}, \boldsymbol{\ell_3}),
\end{align}
where $\mathcal{G}^{\ell_1 \ell_2 \ell_3}_{m_1m_2m_3}$ is the Gaunt integral:
\begin{align}
    \label{eq:gaunt}
    \mathcal{G}^{\ell_1 \ell_2 \ell_3}_{m_1m_2m_3} &= \sqrt{\frac{(2\ell_1+1)(2\ell_2+1)(2\ell_3+1)}{4\pi}} \nonumber\\
    &\times \left(\begin{matrix}
\ell_1 & \ell_2 & \ell_3\\
0 & 0 & 0
\end{matrix}\right)\left(\begin{matrix}
\ell_1 & \ell_2 & \ell_3\\
m_1 & m_2 & m_3
\end{matrix}\right),
\end{align}
in which the final matrix on the R.H.S. is the Wigner 3$j$-symbol.

However, equation (\ref{eq:bidef}) is highly challenging computationally, due to the multiple nested-integrals that need to be calculated. Fortunately, this calculation can be simplified by recognizing that, given that the convergence is a projection of the density contrast, the convergence bispectrum is a projection of the matter bispectrum, $B_{\delta\delta\delta}$, and the matter bispectrum is separable. This means that it can be expressed as the linear sum of products of functions of momenta:
\begin{align}
    \label{eq:matbisep}
    B_{\delta\delta\delta}(\boldsymbol{k_1}, \boldsymbol{k_2}, \boldsymbol{k_3}; \chi_1, \chi_2, \chi_3) &= \sum_{n_1, n_2, n_3} f_{1; n_1}(k_1, \chi_1)\nonumber\\
    &\times f_{2; n_2}(k_2, \chi_2)\nonumber\\
    &\times f_{3; n_3}(k_3, \chi_3),
\end{align}
where $n_1, n_2, n_3$ are power-law indices in their respective functions. For a review of why this holds true, see \cite{LeeDvorkin}. Now, the convergence bispectrum can be expressed as:
\begin{align}
    \label{eq:convbisep}
    B_{ijq}^{\kappa\kappa\kappa}(\boldsymbol{\ell_1}, \boldsymbol{\ell_2}, \boldsymbol{\ell_3}) &= \frac{1}{(2\pi^2)^3}\int_0^{\chi_{\rm lim}} {\rm d}r\,r^2[I^{(1,n_1)}_{\ell_1; i}(r)\nonumber\\
    &\times I^{(2,n_2)}_{\ell_2; j}(r)I^{(3,n_3)}_{\ell_3; q}(r)+ {\rm perms.}],
\end{align}
within which:
\begin{align}
    \label{eq:Ibigdef}
    I^{(a,n_a)}_{\ell_n}(r) &= 4\pi \int_0^{\chi_{\rm lim}} {\rm d}\chi\, W(\chi) \int_0^\infty {\rm d}k\, j_\ell(k\chi)j_\ell(kr)\nonumber\\
    &\times k^2 f_{a; n_a}(k, \chi).
\end{align}
Spherical Bessel functions are highly oscillatory, making the integrals in equation (\ref{eq:Ibigdef}) a significant computational challenge. To bypass this, we can realize that the integral in $k$ will peak when $\chi\simeq r$, and replace the $k$-integral with a Dirac delta function, $\delta_{\rm D}$ :
\begin{align}
    \label{eq:Ilimb}
    I^{(a,n_a)}_{\ell_n}(r) &\approx 4\pi \int_0^{\chi_{\rm lim}} {\rm d}\chi\, W(\chi)\, \frac{\pi}{2r^2}\nonumber\\
    &\times f_{a; n_a}(k=\ell/r, \chi)\delta_{\rm D}(\chi-r)\nonumber\\
    &\approx \frac{2\pi^2}{r^2}W(r)\,f_{a; n_a}(k=\ell/r, r).
\end{align}
This is the application of the Limber approximation.

An additional complication is that the analytic form of the matter bispectrum is not well known. Using second-order perturbation theory (2PT) yields a simple expression for this quantity \cite{Frypap}:
\begin{align}
    \label{eq:Blin}
    B_{\delta\delta\delta}(\boldsymbol{k_1}, \boldsymbol{k_2}, \boldsymbol{k_3}; \chi_1, \chi_2, \chi_3) &= 2F_2\,(\boldsymbol{k_1},\boldsymbol{k_2})\nonumber\\
    &\times P_{\delta\delta}^{\rm lin}(k_1, \chi_1)\nonumber\\
    &\times P_{\delta\delta}^{\rm lin}(k_2, \chi_2) \nonumber\\
    &+ \text{cyc. perms.}, 
\end{align}
where $P_{\delta\delta}^{\rm lin}$ is the linear matter power spectrum, and:
\begin{align}
\label{eq:Fnorm}
        F_2(\boldsymbol{k_1},\boldsymbol{k_2}) &= \frac{5}{7}+ \frac{1}{2}\frac{\boldsymbol{k_1}\cdot\boldsymbol{k_2}}{k_1k_2}\,\bigg(\frac{k_1}{k_2}+\frac{k_2}{k_1}\bigg) \nonumber\\
        &+\frac{2}{7}\,\bigg(\frac{\boldsymbol{k_1}\cdot\boldsymbol{k_2}}{k_1k_2}\bigg)^2.
\end{align}
These expressions are valid in the linear regime, where $\ell_1, \ell_2, \ell_3 < 100$ (see e.g. \cite{Bispeclin}). In order to be able to accurately capture the behaviour of the bispectrum beyond this range, we must obtain fitting formulae from N-body simulations \cite{ScocCouch, Gilmarin, bihalofit}. In this work, we use the fitting formula of \cite{ScocCouch}, to allow consistent comparison with the results of \cite{Deshpap}. Then, in equation (\ref{eq:Blin}), $P_{\delta\delta}^{\rm lin}$ is replaced by $P_{\delta\delta}$, and $F_2$ is replaced by:
\begin{align}
\label{eq:Feff}
        F_2^{\rm eff}(\boldsymbol{k_1},\boldsymbol{k_2}) &= \frac{5}{7}\,a(n_{\rm s},k_1)\,a(n_{\rm s},k_2) \nonumber\\
        &+ \frac{1}{2}\frac{\boldsymbol{k_1}\cdot\boldsymbol{k_2}}{k_1k_2}\,\bigg(\frac{k_1}{k_2}+\frac{k_2}{k_1}\bigg)\,b(n_{\rm s},k_1)\,b(n_{\rm s},k_2) \nonumber\\
        &+\frac{2}{7}\,\bigg(\frac{\boldsymbol{k_1}\cdot\boldsymbol{k_2}}{k_1k_2}\bigg)^2c(n_{\rm s},k_1)\,c(n_{\rm s},k_2),
\end{align}
where $n_{\rm s}$ is the scalar spectral index, and $a,b$, and $c$ are fitting functions detailed in \cite{ScocCouch}.

\subsection{\label{subsec:fishandbi}Fisher Matrices and Biases}

We estimate the biases in cosmological parameters that will be inferred from a \emph{Euclid}-like survey due to neglected systematic effects, by using the Fisher matrix formalism \citep{Tegmark97}. The Fisher matrix is defined as the expectation of the Hessian of the likelihood:
\begin{equation}
    \label{eq:fishbasic}
    F_{\tau\zeta} = \bigg\langle\frac{-\partial^2 \ln L}{\partial\theta_\tau\partial\theta_\zeta}\bigg\rangle,
\end{equation}
where $L$ is the likelihood of the parameters given the data, and $\tau$ and $\zeta$ refer to parameters of interest, $\theta_\tau$ and $\theta_\zeta$. Assuming a Gaussian likelihood, the Fisher matrix can be rewritten in terms of only the covariance of the data, $\boldsymbol{C}$, and the mean of the data vector, $\boldsymbol{\mu}$:
\begin{align}
    \label{eq:fishgauss}
    F_{\tau\zeta} &= \frac{1}{2}\:{\rm tr}\,\bigg[\frac{\partial\boldsymbol{C}}{\partial\theta_\tau}\boldsymbol{C}^{-1}\frac{\partial\boldsymbol{C}}{\partial\theta_\zeta}\boldsymbol{C}^{-1}\bigg] \nonumber\\
    &+ \sum_{pq}\frac{\partial\mu_p}{\partial\theta_\tau}(\boldsymbol{C}^{-1})_{pq}\frac{\partial\mu_q}{\partial\theta_\zeta},
\end{align}
where the summations over $p$ and $q$ are summations over the variables in the data vector. In the case of cosmic shear, we can take our signal to be the mean of the power spectrum, so the first term in equation (\ref{eq:fishgauss}) vanishes. For cosmic shear we can express the covariance, under the assumption of Gaussianity, as:
\begin{align}
    \label{eq:gauscov}
    {\rm Cov}\left[C^{\epsilon\epsilon}_{\ell;ij},C^{\epsilon\epsilon}_{\ell^\prime;mn}\right] &=
\frac{C^{\epsilon\epsilon}_{\ell;i m}C^{\epsilon\epsilon}_{\ell^\prime;jn}+C^{\epsilon\epsilon}_{\ell;in}C^{\
\epsilon\epsilon}_{\ell^\prime;jm}}{(2 \ell + 1) f_{\rm sky} \Delta \ell}\delta^{\rm K}_{\ell\ell^\prime}
\end{align}
where $\delta^{\rm K}$ is the Kronecker delta, $\Delta\ell$ is the bandwidth of $\ell$-modes sampled, and $f_{\rm sky}$ is the fraction of the sky surveyed.
The resulting Fisher matrix can then be expressed as:
\begin{align}
    \label{eq:fishGauss}
    F_{\tau\zeta} &= \sum_{\ell=\ell_{\rm min}}^{\ell_{\rm max}}\sum_{ij,mn}
             \frac{\partial C^{\epsilon\epsilon}_{\ell;ij}}{\partial \theta_{\tau}} {\rm Cov}^{-1}\left[C^{\
\epsilon\epsilon}_{\ell;ij},C^{\epsilon\epsilon}_{\ell';mn}\right]\nonumber\\
&\times\frac{\partial C^{\epsilon\epsilon}_{\ell';mn}}{\partial \theta_{\zeta}},
\end{align}
where $(\ell_{\rm min}, \ell_{\rm max})$ are the minimum and maximum angular wavenumbers used, the sum is over the $\ell$-blocks, and where $(i,j)$ and $(m,n)$ are redshift bin pairs. 

The Fisher matrix can then be used to calculate the expected uncertainties on our parameters, $\sigma_\tau$, using the relation:
\begin{equation}
    \label{eq:sigma}
    \sigma_\tau = \sqrt{({F^{-1}})_{\tau\tau}}.
\end{equation}

In the presence of a neglected systematic effect in the signal, the Fisher matrix formalism can be adapted to measure how biased the inferred cosmological parameter values will be \citep{Fishbias}:
\begin{align}
    \label{eq:bias}
    b(\theta_\tau) &= \sum_\zeta{(F^{-1})}_{\tau\zeta}\: B_\zeta.
\end{align}
Then $B_\zeta$ is given by:
\begin{align}
    \label{eq:sscbias}
    B_{\zeta} &= \sum_{\ell=\ell_{\rm min}}^{\ell_{\rm max}}\sum_{ij,mn}\delta C_{\ell;ij}\,
{\rm Cov}^{-1}\left[C^{\
\epsilon\epsilon}_{\ell;ij},C^{\epsilon\epsilon}_{\ell';mn}\right]\nonumber\\
&\times\frac{\partial C_{\ell';mn}}{\partial \zeta},
\end{align}
where $\delta C_{\ell;ij}$ is the value of the systematic effect for bins $(i,j)$. In this work this systematic effect is the difference between the post-Limber and Limber approximated reduced shear and magnification bias.

\section{\label{sec:method}Methodology}

To study the impact of relaxing the Limber approximation in the bispectra and corrections terms of a \emph{Euclid}-like survey, we adopt the modelling specifications of \cite{ISTFpap}. Accordingly, we compute our chosen quantities for $\ell$-modes up to 5000, as these are required for such a survey to meet its precision goals with cosmic shear. We also take the fraction of sky covered by the survey, $f_{\rm sky}$ to equal 0.36, and the intrinsic variance of observed ellipticities to have two components, each with a value of 0.21, so that the intrinsic ellipticity root-mean-square value $\sigma_\epsilon = \sqrt{2}\times0.21 \approx 0.3$.

A \emph{Euclid}-like survey would be expected to have ten equi-populated redshift bins, covering the range 0 -- 2.5. However, in this work, we only compute the bispectra and correction terms for the auto-correlation of four redshift bins: $[0.001, 0.418]$, $[0.678, 0.789]$, $[1.019, 1.155]$, and $[1.576, 2.50]$. These serve to sufficiently illustrate the impact of the Limber approximation across the survey's redshift range, while avoiding the significant computational expense of computing the 55 total bin combinations.

Next, for photometric redshift estimates, we define the galaxy distributions in our tomographic bins as:
\begin{equation}
    \label{eq:ncfht}
    {\mathcal N}_i(z) = \frac{\int_{z_i^-}^{z_i^+}{\rm d}z_{\rm p}\,\mathfrak{n}(z)p_{\rm ph}(z_{\rm p}|z)}{\int_{z_{\rm min}}^{z_{\rm max}}{\rm d}z\int_{z_i^-}^{z_i^+}{\rm d}z_{\rm p}\,\mathfrak{n}(z)p_{\rm ph}(z_{\rm p}|z)},
\end{equation}
where $z_{\rm p}$ is measured photometric redshift, $z_i^-$ and $z_i^+$ are edges of the $i$-th redshift bin, $z_{\rm min}$ and $z_{\rm max}$ define the range of redshifts covered by the survey, and $\mathfrak{n}(z)$ is the true distribution of galaxies with redshift, $z$, defined as \cite{EuclidRB}:
\begin{equation}
    \label{eq:ntrue}
    \mathfrak{n}(z) \propto \bigg(\frac{z}{z_0}\bigg)^2\,{\rm exp}\bigg[-\bigg(\frac{z}{z_0}\bigg)^{3/2}\bigg],
\end{equation}
where $z_0=z_{\rm m}/\sqrt{2}$, with $z_{\rm m}=0.9$ as the median redshift of the survey. Additionally, the function $p_{\rm ph}(z_{\rm p}|z)$ describes the probability that a galaxy at redshift $z$ is measured to have a redshift $z_{\rm p}$, and takes the parameterisation:
\begin{align}
\label{eq:pphot}
        p_{\rm ph}(z_{\rm p}|z) &= \frac{1-f_{\rm out}}{\sqrt{2\pi}\sigma_{\rm b}(1+z)}\,{\rm exp}\Bigg\{-\frac{1}{2}\bigg[\frac{z-c_{\rm b}z_{\rm p}-z_{\rm b}}{\sigma_{\rm b}(1+z)}\bigg]^2\Bigg\} \nonumber\\
        &+ \frac{f_{\rm out}}{\sqrt{2\pi}\sigma_{\rm o}(1+z)}\nonumber\\
        &\times{\rm exp}\Bigg\{-\frac{1}{2}\bigg[\frac{z-c_{\rm o}z_{\rm p}-z_{\rm o}}{\sigma_{\rm o}(1+z)}\bigg]^2\Bigg\}.
\end{align}
Here, the first term describes the multiplicative and additive bias in redshift determination for the fraction of sources with a well measured redshift, whereas the second term accounts for the effect of a fraction of catastrophic outliers, $f_{\rm out}$. The values of the parameters used in equation (\ref{eq:pphot}) are stated in Table \ref{tab:phphotparams}.
\begin{table}[b]
    \centering
    \caption{Parameter values used to define the probability distribution function of the photometric redshift distribution of sources, in Eq. (\ref{eq:pphot}).}
    \begin{tabular}{c c}
    \hline\hline
    Model Parameter& Fiducial Value\\
    \hline
    $c_{\rm b}$ & 1.0\\
    $z_{\rm b}$ & 0.0\\
    $\sigma_{\rm b}$ & 0.05\\
    $c_{\rm o}$ & 1.0\\
    $z_{\rm o}$ & 0.1\\
    $\sigma_{\rm o}$ & 0.05\\
    $f_{\rm out}$ & 0.1\\
    \hline\hline
    \end{tabular}
    \label{tab:phphotparams}
\end{table}
Then, $n_i(\chi) = {\mathcal N}_i(z){\rm d}z/{\rm d}\chi$.

For our fiducial cosmology, we choose $w_0w_a$CDM, which has the following parameters: the present-day matter density parameter $\Omega_{\rm m}$, the present-day baryonic matter density parameter $\Omega_{\rm b}$, the Hubble parameter $h=H_0/100$km\:s$^{-1}$Mpc$^{-1}$, the spectral index $n_{\rm s}$, the RMS value of density fluctuations on 8 $h^{-1}$Mpc scales $\sigma_8$, the present-day dark energy density parameter $\Omega_{\rm DE}$, the present-day value of the dark energy equation of state $w_0$, the high redshift value of the dark energy equation of state $w_a$, and massive neutrinos with sum of masses $\sum m_\nu\ne 0$. This model of cosmology allows for a time-varying dark energy equation-of-state. The values for these parameters are chosen to match those of \cite{ISTFpap} and \cite{Deshpap}, and are stated in Table \ref{tab:cosmology}. 
\begin{table}[t]
\centering
\caption{Fiducial values of $w_0w_a$CDM cosmological parameters for which the bispectra, and reduced shear and magnification bias corrections are calculated. These values were selected in accordance with \citep{ISTFpap} and \cite{Deshpap} to facilitate consistent comparisons.}
\label{tab:cosmology}
\begin{tabular}{c c}
\hline\hline
Cosmological Parameter & Fiducial Value\\
\hline
$\Omega_{\rm m}$ & 0.32 \\
$\Omega_{\rm b}$ & 0.05 \\
$h$ & 0.67 \\
$n_{\rm s}$ & 0.96 \\
$\sigma_8$ & 0.816 \\
$\sum m_\nu$ (eV) & 0.06 \\
$\Omega_{\rm DE}$ & 0.68 \\
$w_0$ & $-1$ \\
$w_a$ & 0  \\
\hline\hline
\end{tabular}
\end{table}

The matter density power spectrum used in our analyses is computed using the publicly available \texttt{CLASS}\footnote{\url{https://class-code.net/}} cosmology package \citep{Classpap}. Within \texttt{CLASS}, we included non-linear corrections to the matter density power spectrum, using the \texttt{Halofit} model \citep{Takahashi12}. To model the IA power spectra, we choose $\mathcal{A}_{\rm IA}=1.72$ and $\mathcal{C}_{\rm IA}=0.0134$ \cite{Deshpap, ISTFpap}. 

Using the discussed modelling specifications, we compute the convergence bispectrum, both with and without making the Limber approximation. We compute the bispectra for equilateral configurations where $\ell_1=\ell_2=\ell_3$, and isosceles configurations where $\ell_1=\ell_2\neq\ell_3$. We present two different isosceles configurations, one where $\ell_3=20$, and another where $\ell_3=100$. The separability of the bispectrum is used to reduce some of the computational complexity of the post-Limber case. The 2PT expression for the bispectrum stated in equation (\ref{eq:Blin}) is valid when $\ell_1, \ell_2, \ell_3 < 100$. Accordingly, it is also true that the bispectrum's individual separated components $I^{a, n_a}_{\ell_n}$, as defined in equation (\ref{eq:Ibigdef}), will match sufficiently well whether only the 2PT expression or the non-linear fitting function expression of equation (\ref{eq:Feff}) is used in their computation, when $\ell_n<100$. Therefore, using the 2PT expression for a particular $I^{a, n_a}_{\ell_n}$ when the corresponding $\ell_n<100$, avoids the laborious numerical integration over the fitting functions of equation (\ref{eq:Feff}) in that case; reducing the overall total computation time.

We then compute the reduced shear and magnification bias corrections. However, the integration over the bispectrum necessitated by these terms is an intractable computation to perform directly for the post-Limber case, given the number of steps in $\ell$-space required. It would take on the order of $\sim$ 50 weeks for just one bin auto-correlation\footnote{For a \texttt{Python} script multi-processed across 100 CPU threads.}. To bypass this hurdle, we first compute the post-Limber and Limber approximated bispectra on a grid of 1331 points in $\ell$-space, with each $\ell_i$ sampled logarithmically in the range $10\leq\ell_i\leq5000$, for each bin. The ratio of these quantities at each point is then taken, and these ratios are interpolated over, using linear 3D interpolation. This gives a function which maps the Limber approximated bispectrum onto the post-Limber case. In computing the post-Limber reduced shear and magnification bias corrections, we calculate the required bispectra as in the Limber approximated case, and use the previously interpolated function to scale these to their post-Limber counterparts.
\begin{table}[t]
    \centering
    \caption{Slope of the luminosity function for studied redshift bins, calculated at the central redshifts of each bin. These are evaluated at the limiting magnitude 24.5. The slopes are determined using finite difference methods with the fitting formula of \cite{JoachimiBridleMB}.} 
    \begin{tabular}{c c c}
    \hline\hline
    Bin $i$ & Central Redshift & Slope $s_i$\\
    \hline
    0.001 - 0.418 & 0.2095 & 0.196\\
    0.678 - 0.789 & 0.7335 & 0.365\\
    1.019 - 1.155 & 1.087 & 0.525\\
    1.576 - 2.50 & 2.038 & 1.089\\
    \hline\hline
    \end{tabular}
    \label{tab:slopesbybin}
\end{table}
To compute the magnification bias correction, we require the slope of the galaxy luminosity function. To calculate this quantity, we use the approach described in \cite{Deshpap}. Accordingly, we use the fitting function for galaxy number density as a function of limiting magnitude derived in Appendix C of \cite{JoachimiBridleMB}, and calculate its slope by using a finite difference method. For each redshift bin, we calculate the slope at the central redshift of that bin. For a \emph{Euclid}-like survey, we take the limiting magnitude to be 24.5 \cite{EuclidRB}. The determined slopes are listed in Table \ref{tab:slopesbybin}.

We compare these corrections to the sample variance of a \emph{Euclid}-like survey. The sample variance from LSS for a weak lensing galaxy survey is given by:
\begin{equation}
    \label{eq:sampvar}
    \delta C_{\ell; ij}^{\rm SV}\,/\,C_{\ell;ij}^{\gamma\gamma} = \sqrt{2}\,[f_{\rm sky}(2\ell +1)]^{-1/2},
\end{equation}
where $f_{\rm sky}$ is the fraction of surveyed \cite{Petespasflat}.

Then, we define a worst-case scenario where the difference between post-Limber and Limber approximated corrections, $ \Delta C_{\ell; ij}\approx 0.01\,C^{\gamma\gamma}_{\ell; ij}$ for all bin combinations and $\ell$-modes. This corresponds to the largest difference seen for our survey specifications, at $\ell\approx5000$ for the auto-correlation of bin $1.56-2.50$. Using equation (\ref{eq:fishgauss}) and equation (\ref{eq:sscbias}), we then calculate the cosmological parameter biases resulting from $\Delta C_{\ell; ij}$. For this calculation, we consider the auto and cross-correlation spectra for all ten redshift bins expected for a \emph{Euclid}-like survey, with bin edges: \{0.001, 0.418, 0.560, 0.678, 0.789, 0.900, 1.019, 1.155, 1.324, 1.576, 2.50\} \cite{Deshpap,ISTFpap}. The Fisher matrix we construct contains the parameters $\Omega_{\rm m}, \Omega_{\rm b}, h, n_{\rm s}, \sigma_8, \Omega_{\rm DE}, w_0, w_a,$ and $\mathcal{A}_{\rm IA}$.

\section{\label{sec:results}Results and Discussion}

Here, we present the effect of relaxing the Limber approximation on the quantities examined. Firstly, we report the impact on the convergence bispectrum in the four studied redshift bins. Then, we do the same for the reduced shear and magnification bias corrections to the angular power spectrum.

\subsection{\label{subsec:resbispec}The Post-Limber Convergence Bispectrum}

\begin{figure}[H]
\centering
\includegraphics[width=0.845\linewidth]{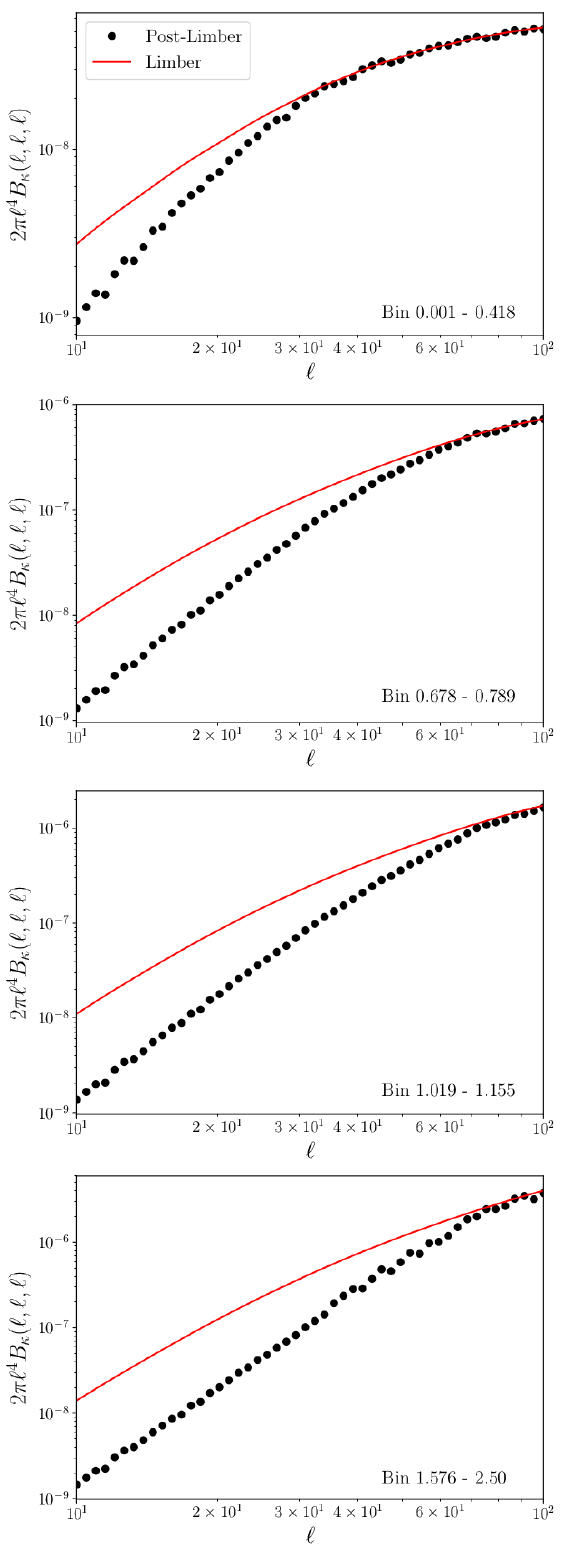}
\caption{Comparison of the equilateral configuration convergence bispectrum with and without making the Limber approximation, for the auto-correlation of four redshift bins across the redshift range of a \emph{Euclid}-like survey. The Limber approximation fails when $\ell<60$, and overestimates the bispectrum. This over-prediction is worse at higher redshifts and lower $\ell$-modes.}
\label{fig:nl_bispec}
\end{figure}

The effect of relaxing the Limber approximation for the equilateral configuration of the convergence bispectrum is shown in Figure (\ref{fig:nl_bispec}), for all of the examined redshift bins. From this, we see that, for all redshift bins, the Limber approximation over-predicts the bispectrum for \makebox[\linewidth][s]{$\ell$-modes below $\ell \sim 60$. Additionally, the over-prediction}
\noindent worsens at lower $\ell$-modes, and for higher redshift bins.

\begin{figure*}[p]
\centering
\includegraphics[width=0.845\linewidth]{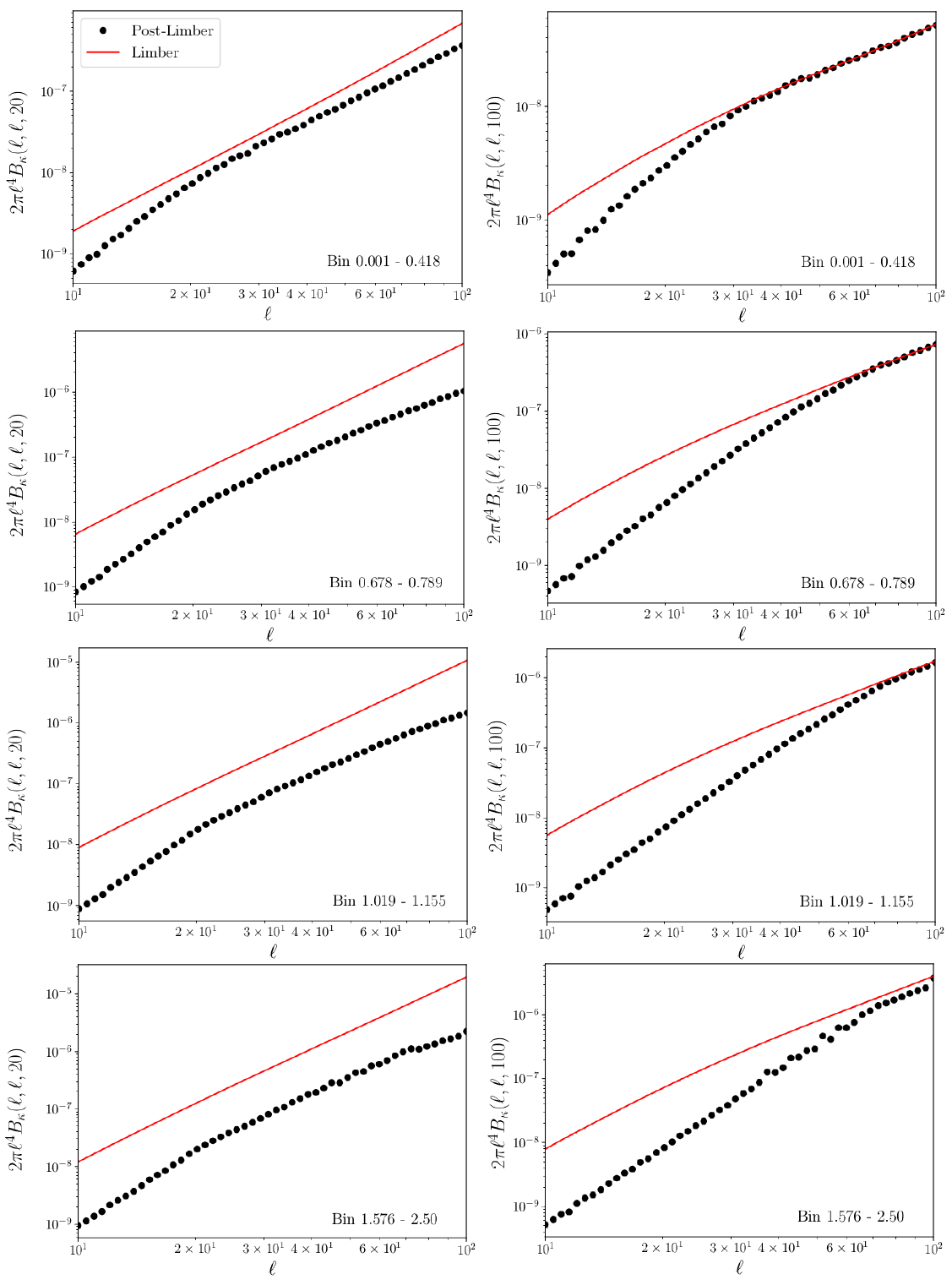}
\caption{Isosceles configuration bispectra with $\ell_3=20$ (\textbf{left column}), and $\ell_3=100$ (\textbf{right column}), for four tomographic bins across the redshift range of a \emph{Euclid}-like survey. The values of the bispectra with and without making the Limber approximation are shown. When $\ell_3=20$, the Limber approximation over-predicts the bispectrum for all values of $\ell_1$ and $\ell_2$. However, when $\ell_3=100$, we see similar behaviour to the equilateral case shown in Figure (\ref{fig:nl_bispec}), in that the Limber approximation only results in over-prediction for $\ell_1=\ell_2<60$. This suggests the Limber approximation fails when any one of the sides of the bispectrum triangle $\ell_i<60$. Otherwise, the trends in both displayed isosceles cases match those seen for the equilateral configuration, with over-prediction worsening at higher redshift, and lower $\ell$.}
\label{fig:nl_biiso}
\end{figure*}

Furthermore, Figure (\ref{fig:nl_biiso}) shows the bispectra on the four bins for two different isosceles configurations. The configurations shown are when $\ell_1=\ell_2$ and $\ell_3=20$, and when $\ell_1=\ell_2$ and $\ell_3=100$. For the former of these cases, we notice that the bispectrum is over-predicted by the Limber approximation for all $\ell$-modes. On the other hand, when in an isosceles configuration with $\ell_3=100$, we see much the same trends as in Figure (\ref{fig:nl_bispec}). This implies that the Limber approximation fails for the convergence bispectrum when any one of its sides $\ell_i < 60$. Similar discrepancies at low $\ell$-modes are seen for both the equilateral and isosceles configurations in \cite{DipakBispec}, where the Limber-approximated theoretical expression for the bispectrum is compared to the bispectrum measured from full-sky simulations.

Accordingly, if $\ell$-modes below 60 are probed, as will be the case for Stage IV experiments, the Limber approximation cannot be used to compute the bispetrum in this regime. This presents a computational challenge, as computing the post-Limber bispectrum is two orders-of-magnitude slower than using the Limber approximation. However, the separability of the bispectrum, discussed in Section \ref{subsec:bispec}, offers a solution. For a given configuration, if one of the sides of the bispectrum $\ell_i < 60$, only the instances of equation (\ref{eq:Ibigdef}) corresponding to that side need to be computed without the Limber approximation. Furthermore, there has recently been great success in using the FFTLog decomposition technique to significantly speed up the the computation of higher-order statistics without the Limber approximation \cite{IASpap, LeeDvorkin}.

\subsection{\label{subsec:rescorrs}The Post-Limber Reduced Shear and Magnification Bias Corrections}

\begin{figure}[H]
\centering
\includegraphics[width=0.875\linewidth]{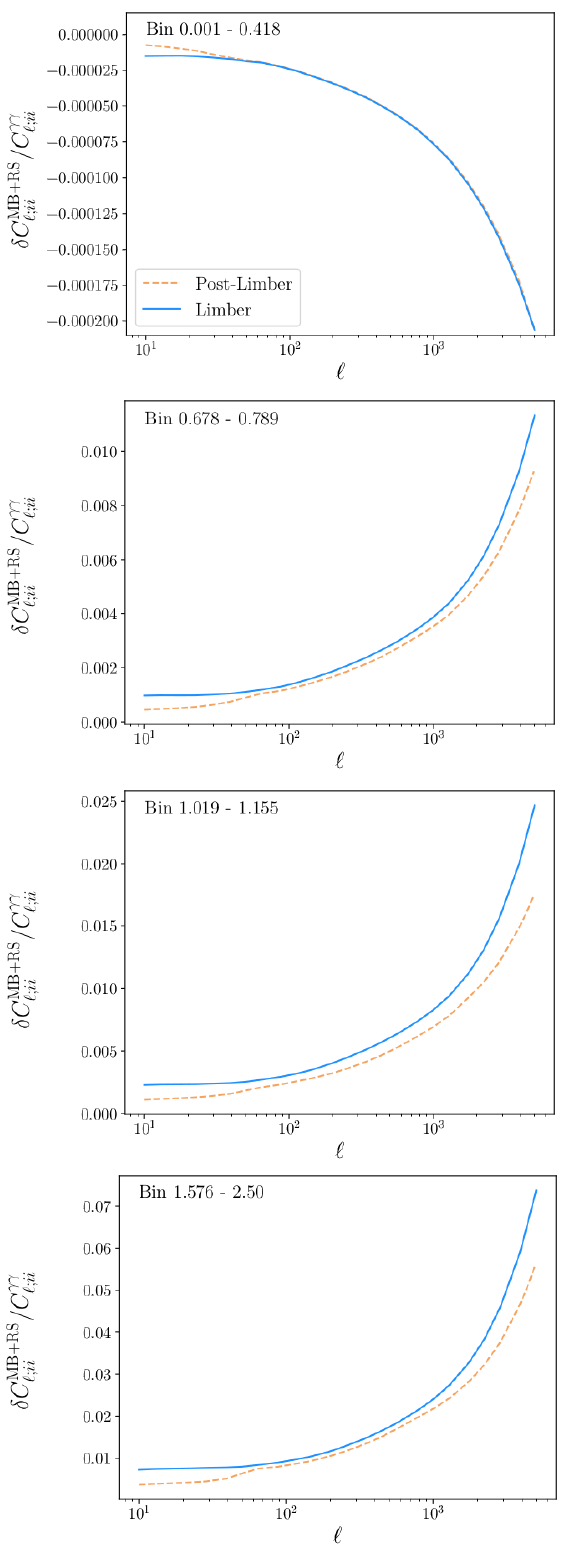}
\caption{Combined reduced shear and magnification bias corrections, with and without making the Limber approximation. Corrections are displayed for the auto-correlation of four bins across the redshift range of a \emph{Euclid}-like survey, 0 -- 2.5. Now, due to mode-mixing, the Limber approximation overestimates the correction terms at all $\ell$-modes. As with the convergence bispectra, the over-prediction worsens at higher redshift.}
\label{fig:dcl_nl}
\end{figure}

Figure (\ref{fig:dcl_nl}) shows the impact of relaxing the Limber approximation on the combined corrections to the cosmic shear angular power spectra for the reduced shear approximation and magnification bias. Now, we see that the magnitude of these corrections is over-estimated slightly throughout the entire probed range when the Limber approximation is made. This is due to the fact that the mathematical forms of these corrections, equation (\ref{eq:dCl}) and equation (\ref{eq:dClMB}), involve integrating over two of the sides of the bispectrum triangle.

\clearpage
\begin{figure}[H]
\centering
\includegraphics[width=1.0\linewidth]{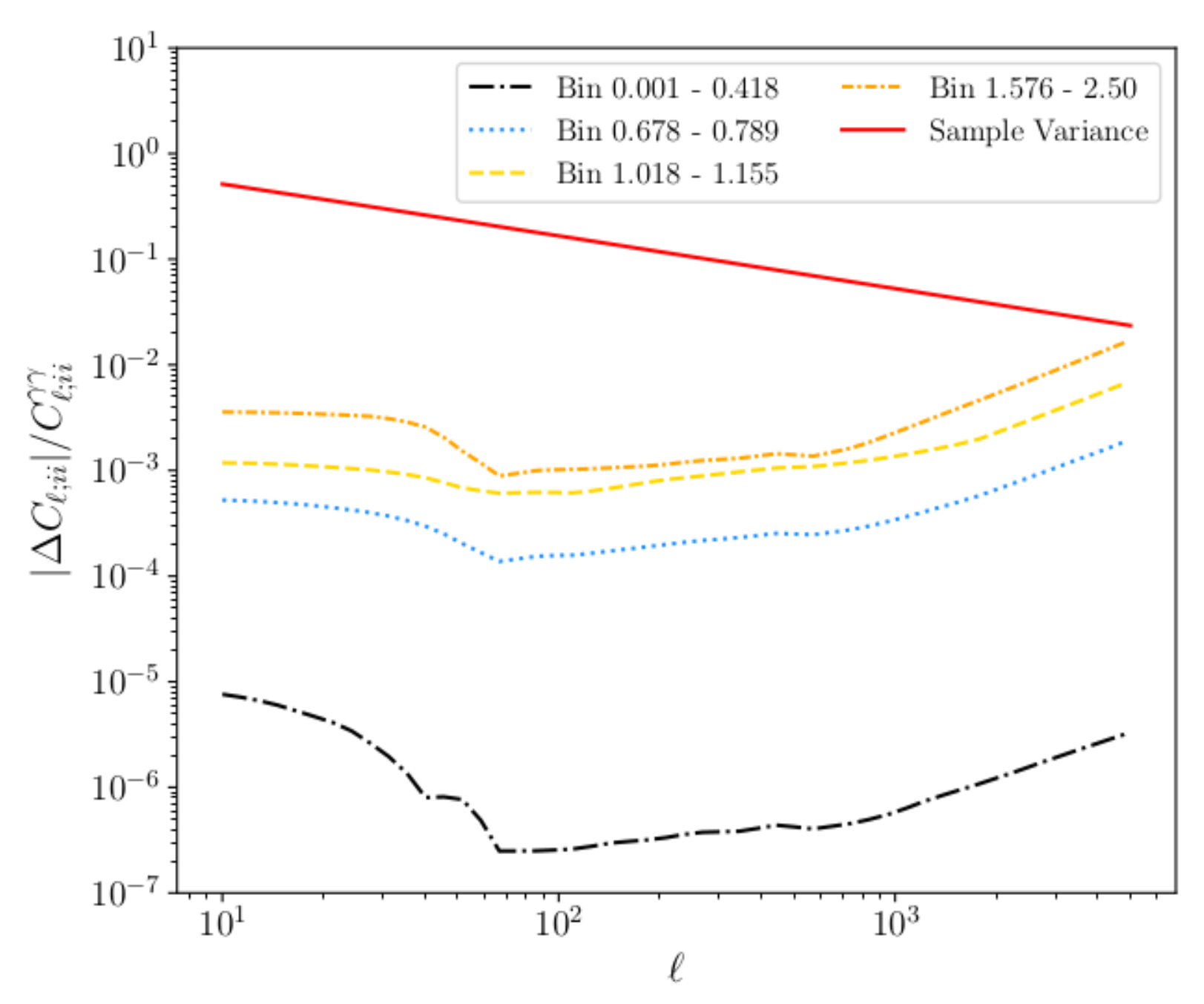}
\caption{Difference between Limber and post-Limber reduced shear and magnification bias corrections, relative to the auto-correlation power spectrum for four bins in redshift range 0 -- 2.5. The sample variance of the galaxy survey is also shown for comparison. The overestimation of the Limber approximation worsens at higher redshifts. However, it is below sample variance across the probed redshift range; meaning that the Limber approximation is sufficient when calculating these correction terms for Stage IV experiments.}
\label{fig:svcomp}
\end{figure}

Accordingly, mode-mixing results in bispectra with at least one $\ell$-mode less than 60 being involved in corrections for all $\ell$ values. We also see that, once again, the over-estimation is worse for the higher redshift bins. This is expected, given that these trends are carried across from the bins' respective bispectra. The correction terms themselves are highest at higher redshift; meaning that they are the dominant contribution to the induced cosmological biases \cite{Deshpap}.

However, for all bins, the difference between the Limber approximated and post-Limber cases is below sample variance, as seen in Figure (\ref{fig:svcomp}). Additionally, the worst-case scenario cosmological parameter biases, when $ \Delta C_{\ell; ij}\approx 0.01\,C^{\gamma\gamma}_{\ell; ij}$, are stated in Table \ref{tab:biases}. Also reproduced here, from \cite{Deshpap}, are the biases if the reduced shear and magnification bias corrections are neglected entirely.

The bias on a parameter is considered significant when it exceeds $0.25\sigma$, as at this point the confidence contours of the parameters with and without the systematic overlap less than $90\%$ \cite{Fishbias}. From Table \ref{tab:biases}, we see that none of the biases are significant. In fact, all but one of the biases have a magnitude less than $0.20\sigma$ which means that the confidence regions of those parameters having neglected the bias have an overlap of more than 95$\%$ with the parameters' confidence regions when the bias is taken into account.

We note that the bias in the inferred value of $w_a$ sits on the threshold of significance. However, for $\ell\lesssim 5000$ and all bin correlations other than the auto-correlation of bin $1.576-2.50$, $\Delta C_{\ell; ij} < 0.01\,C^{\gamma\gamma}_{\ell; ij}$. Given that these modes and bins will make up the majority of observations for a \emph{Euclid}-like survey, we can safely conclude that the cosmological biases induced from neglecting the reduced shear and magnification bias corrections will not be significantly altered by whether they make the Limber approximation or not.

Furthermore, biases from the difference between the post-Limber and Limber approximated corrections are significantly smaller in magnitude than those resulting from simply neglecting the Limber approximated corrections entirely. Accordingly, these correction terms can be safely calculated under the Limber approximation for Stage IV experiments.
\begin{table}[t]
\centering
\caption{Worst-case scenario biases in $w_0w_a$CDM cosmological parameters from the difference in the post-Limber and Limber approximated (labelled `PL-L') reduced shear and magnification bias corrections, relative to the predicted $1\sigma$ uncertainty on those parameters \cite{ISTFpap} for a \emph{Euclid}-like survey. The biases resulting from neglecting the reduced shear and magnification bias corrections altogether are also reproduced from \cite{Deshpap}, in the column labelled `AD19'.}
\label{tab:biases}
\begin{tabular}{c c c}
\hline\hline
Cosmological & Worst-case PL-L & AD19\\
Parameter & Bias/$\sigma$ & Bias/$\sigma$\\
\hline
$\Omega_{\rm m}$ & 0.073 & $-$0.53 \\
$\Omega_{\rm b}$ & 0.065 & $-$0.20\\
$h$ & 0.090 & 0.040\\
$n_{\rm s}$ & $-$0.16 & $-$0.34\\
$\sigma_8$ & $-$0.020 & 0.43\\
$\Omega_{\rm DE}$ & 0.13 & 1.36\\
$w_0$ & $-0.18$ & $-$0.68\\
$w_a$ & 0.25 & 1.21\\
\hline\hline
\end{tabular}
\end{table}
\section{\label{sec:conclusions}Conclusions}

Within this work, we have considered how the Limber approximation will affect the convergence bispectrum calculated for Stage IV weak lensing experiments. Additionally, we also calculated the resulting impact on the reduced shear and magnification bias corrections to the angular power spectrum, as these quantities depend on the bispectrum. We found that the Limber approximation significantly over-predicts the bispectrum at $\ell$-modes below 60, throughout the redshift range of a \emph{Euclid}-like survey.

Furthermore, we found this discrepancy worsens at higher redshifts and lower $\ell$ scales. Accordingly, we found that the reduced shear and magnification bias corrections are also over-estimated by the Limber approximation, although the difference was well below the sample variance of a Stage IV weak lensing experiment. Finally, we calculated the worst-case scenario cosmological parameter biases that result from the difference between the post-Limber and Limber approximated corrections. These were found not to be significant. Hence, we conclude that the Limber approximation is sufficient for these terms at this level of precision.

\begin{acknowledgments}
ACD and TDK wish to acknowledge the support of the Royal Society. We thank the Euclid Consortium Editorial Board for guidance on the submission of the article. 
\end{acknowledgments}

\bibliography{apssamp}% Produces the bibliography via BibTeX.

\end{document}